\documentclass[showpacs,preprintnumbers,amsmath,amssymb]{revtex4}

\usepackage{amsmath}
\usepackage{graphicx}
\usepackage{epsf}
\usepackage{psfrag}
\usepackage{epsfig}
\usepackage{graphics}

\begin{document}

\newcommand{\be}{\begin{equation}}
\newcommand{\ee}{\end{equation}}
\newcommand{\bq}{\begin{eqnarray}}
\newcommand{\eq}{\end{eqnarray}}
\newcommand{\dt}{\frac{d^3k}{(2 \pi)^3}}
\newcommand{\dtp}{\frac{d^3p}{(2 \pi)^3}}
\newcommand{\kbruto}{\hbox{$k \!\!\!{\slash}$}}
\newcommand{\pbruto}{\hbox{$p \!\!\!{\slash}$}}
\newcommand{\qbruto}{\hbox{$q \!\!\!{\slash}$}}
\newcommand{\lbruto}{\hbox{$l \!\!\!{\slash}$}}
\newcommand{\bbruto}{\hbox{$b \!\!\!{\slash}$}}
\newcommand{\parbruto}{\hbox{$\partial \!\!\!{\slash}$}}
\newcommand{\Abruto}{\hbox{$A \!\!\!{\slash}$}}

\title{{\bf  Gauge invariance and the CPT and Lorentz violating induced Chern-Simons-like term in extended QED}}

\date{\today}

\author{A. P. Ba\^eta Scarpelli$^{(a)}$} \email[]{scarp1@des.cefetmg.br}
\author{Marcos Sampaio$^{(b)}$}\email[]{msampaio@fisica.ufmg.br}
\author{M. C. Nemes$^{(b)}$}\email[]{carolina@fisica.ufmg.br}
\author{B. Hiller$^{(c)}$} \email[]{brigitte@teor.fis.uc.pt}

\affiliation{(a) Centro Federal de Educa\c{c}\~ao Tecnol\'ogica de Minas Gerais \\
Avenida Amazonas, 7675 - 30510-000 - Nova Gameleira - Belo Horizonte
-MG - Brazil}
\affiliation{(b) Universidade Federal de Minas Gerais - Departamento de F\'{\i}sica - ICEx \\
P.O. BOX 702, 30.161-970, Belo Horizonte MG -
Brazil}
\affiliation{(c) Departamento de Fisica, Universidade de Coimbra, P-3004516 Coimbra, Portugal }

\begin{abstract}

\noindent
The radiative induction of the CPT and Lorentz violating Chern-Simons (CS)
term is reassessed. The massless and massive models are studied. Special
attention is given to the preservation of gauge symmetry at higher
orders in the background vector $b_\mu$ when radiative
corrections are considered. Both the study of the odd and even parity sectors of the complete vacuum
polarization tensor at one-loop order and a non-perturbative analysis show
that this symmetry must be preserved by the quantum corrections.
As a complement we obtain that transversality of the polarization tensor does not
fix the value of the coefficient of the induced CS term.

\end{abstract}


\maketitle

\section{Introduction}
Symmetries are the cornerstones when one intends to systematize the study of any theory.
Lorentz and CPT invariances have supreme importance in the elaboration of modern Quantum
Field Theory models. However, in the last decade the possibility of violation of these symmetries
has been vastly investigated \cite{2}-\cite{ultlv}. A standard model description, where
possible violations of such invariances are considered, was developed by Colladay and Kosteleck\'y
\cite{5}, \cite{7} and by Coleman and Glashow \cite{6}, \cite{9}. One of the most discussed terms that incorporates
these features has the Chern-Simons (CS) form
\be
\Sigma_{CS}=-\frac 14\int d^4x\, c_\mu A_\nu F_{\alpha \beta} \epsilon^{\mu \nu \alpha \beta},
\ee
in which $c_\mu$ is a constant four-vector that selects a space-time direction. Therefore, it gives rise to
an optical activity of the vacuum. Although astrophysical results put very stringent limits on the magnitude of
$c_\mu$ \cite{2}, \cite{4}, many controversies have emerged from the discussion whether such term
could be generated by means of radiative corrections from the fermionic sector with the inclusion of the
CPT and Lorentz violating axial term, $b_\mu \bar\psi \gamma^\mu \gamma^5 \psi$, $b_\mu$ being a constant
background vector. The main point concerns the regularization dependence of
this term.
As discussed in many of these papers, the finite radiative correction $\Delta c_\mu$ comes from the
cancellation of divergences and, therefore, is in principle regularization dependent. Symmetries are
invoked to argue against or in favor of the generation of such a term.

Other discussions focus on the possibility of considering a source for the $b_\mu$ field  and in
stating gauge invariance in a weak way, which means gauge invariance of the action and not necessarily of the Lagrangian density.

In this paper, we reassess the discussion on the radiative generation of the CS-like term. We give particular attention
to the the possibility of gauge symmetry violation coming from quantum corrections. We also analyze if
gauge invariance can impose some constraint on the coefficient of the CS-like term. The paper is divided as follows:
in section II, the massless model is considered. It is shown at one-loop
order that the even parity piece of the polarization tensor
has the structure of simple spinorial QED, on arguments of assigning the
constant background field to arbitrary rooting of the internal momentum
dependence. The implicit regularization (IR) technique allows to pin down
the exact form of surface terms and its a priori arbitrary parameterized values. A
weak constraint, i.e. a relation among two of the parameters, instead of
fixing both values independently, is seen to be sufficient to ensure
gauge invariance. The odd parity term is shown not to deliver further constraints
on this relation, which leads finally to the indeterminacy of the
coefficient of the CS term. This result is compared to the chirally rotated
theory and the Jacobian of the transformation. In section III, we carry out
the analysis of the massive model, expanding the exact propagator to second
order in the constant background field. Again, a careful study of the surface
terms with the IR method leaves the CS-like term coefficient undetermined.
We summarize our results in section IV.

\section{The massless model}
Concerning the modified QED which includes the axial term in the fermionic sector, some papers
were devoted to discuss the gauge invariance of the model \cite{altschul1}, \cite{altschul2}, \cite{bonneau1},
\cite{bonneau2}. In ref. \cite{altschul1}, B. Altschul has analyzed the massless model and argued that
gauge symmetry is violated at $b^2$ order of the vacuum polarization tensor. In a further work \cite{altschul2}, the author has shown that
an adequate Pauli-Villars regulator is compatible with the gauge invariance of the model.

We would like to show here that the gauge invariance of the action enforces a certain relation among
the a priori arbitrary coefficients which parameterize surface terms. This can be achieved by
considering the whole amplitude. Nevertheless, this is not sufficient to fix unambiguously the coefficient of the radiatively generated CS term.

We begin our reasoning by analyzing the massless case, for which the fermion action is
given by
\be
\Sigma_\psi = \int d^4x \,\, \bar \psi (i \parbruto - e \Abruto - \bbruto \gamma_5) \psi.
\ee
It is instructive to discuss first a non-perturbative calculation (in $b$ and in the coupling constant) of the induced
CS-type term. This calculation has been performed by J.-M. Chung in \cite{chung}.
By making the chiral transformation,
\be
\psi \to e^{-i \gamma_5 b \cdot x}\psi \,\, , \,\, \bar \psi \to \bar \psi e^{-i \gamma_5 b \cdot x},
\ee
we can eliminate the $b_\mu$ vector from the classical action. Nevertheless, at the quantum level, the measure of the generating functional
acquires a factor given by the Jacobian \cite{fujikawa},
\be
J[b_\mu,A_\mu]= exp\left\{ -i\int d^4x \,\, (b \cdot x) {\cal A}[A_\mu](x) \right\},
\ee
with
\be
{\cal A}[A_\mu](x)=\frac {1}{16 \pi^2}\epsilon^{\mu \nu \alpha \beta}F_{\mu \nu}F_{\alpha \beta}.
\ee
We can write
\bq
&&J[b_\mu,A_\mu]= \nonumber \\
&& exp\left\{ -i\int d^4x \,\, \frac{1}{4\pi^2} (b \cdot x)\epsilon^{\mu \nu \alpha \beta}
\partial_\mu A_\nu \partial_\alpha A_\beta \right\},
\eq
which after an integration by parts turns out to be
\be
J[b_\mu,A_\mu]= exp\left\{ i\int d^4x \,\, \frac{1}{4\pi^2} b_\mu\epsilon^{\mu \nu \alpha \beta}
A_\nu \partial_\alpha A_\beta \right\}.
\ee
We see that after the chiral transformation the axial term disappears from the fermionic sector.
As a result the QED Lagrangian is obtained, together with a Jacobian which is taken into account when
quantum corrections are calculated.
Therefore the non-massive model with Lorentz and CPT violation in
the fermionic sector is equivalent to the non-massive QED, if a CS-type term coming from the Jacobian is
added to the radiative correction of the photon self-energy. As a consequence, since the latter model
is gauge invariant, so must be the original one. We will comment at the end of this section on
the coefficient of the CS term.

We now carry out an one-loop analysis for the massless case, following ref. \cite{altschul1}. If the fermion is
non-massive, its propagator can be decomposed as
\be
\frac{i}{\kbruto -\bbruto \gamma_5} = \frac{i}{\kbruto-\bbruto}P_L+ \frac{i}{\kbruto+\bbruto}P_R,
\label{proj}
\ee
where we are using the chiral projectors
\be
P_{R,L}=\frac {1\pm \gamma_5}{2}.
\ee
It is now simple to analyze the full vacuum polarization tensor and achieve some conclusions with respect to
the gauge invariance of the theory and the generation of the CS term. The amplitude is easily written as
\be
\Pi^{\mu \nu}=\frac 12 \left\{ \Pi^{\mu \nu}_+ + \Pi^{\mu \nu}_- + \Pi^{\mu \nu}_{5+}+\Pi^{\mu \nu}_{5-}\right\},
\ee
with
\be
\Pi^{\mu \nu}_{\pm} = \int_k^\Lambda \mbox{tr}\left\{ \frac{\gamma^\mu(\kbruto \pm \bbruto) \gamma^\nu
(\kbruto+ \pbruto \pm \bbruto)}{(k \pm b)^2(k+p \pm b)^2} \right\}
\ee
and
\be
\Pi^{\mu \nu}_{5\pm} = \pm \int_k^\Lambda \mbox{tr}\left\{ \frac{\gamma^\mu(\kbruto \pm \bbruto) \gamma^\nu
(\kbruto+ \pbruto \pm \bbruto)\gamma_5}{(k \pm b)^2(k+p \pm b)^2} \right\},
\ee
where $\int_k \equiv \int \frac{d^4k}{(2 \pi)^4}$ and the superscript $\Lambda$ is used to indicate that some regularization
has been applied. Note that $\Pi^{\mu \nu}_{\pm}$ is simply the vacuum polarization tensor of the simple spinorial QED evaluated
with an arbitrary momentum distribution in the internal lines, $k+k_1$ and $k+k_2$, with $p=k_2-k_1$, for
which $k_1=\pm b$ and $k_2=p\pm b$. Any regularization scheme that respects gauge invariance will
return an answer which is transverse and depends only on $p=k_2-k_1$. In ref. \cite{scarpetal}, this amplitude has been
calculated by means of Implicit Regularization \cite{ir} with arbitrary $k_1$ and $k_2$. In the massless limit, it gives
\bq
&&\Pi_{\pm}^{\mu \nu}=\Pi(p^2)(p^\mu p^\nu-p^2 g^{\mu \nu}) \nonumber \\
&&- 4\left( \alpha_1 g^{\mu\nu}-(k_1^2+k_2^2)\alpha_2 g^{\mu
\nu} \right. \nonumber \\
&& \left.
+\frac{1}{3}(k_{1\alpha}k_{1\beta}+k_{2\alpha}k_{2\beta}
+k_{1\alpha}k_{2\beta})
\alpha_3 g^{\{\mu \nu \alpha \beta\}} \right. \nonumber \\
&& \left. -(k_1+k_2)^{\mu}(k_1+k_2)^{\nu}\alpha_2  \right).
\label{QED}
\eq
In the equation above, $\Pi(p^2)$ includes the divergent part. The quantity $g^{\{\mu \nu \alpha \beta\}}$
denotes the symmetrized product of two $g^{\mu\nu}$. Now, the momentum routing dependent
terms, which cause violation of gauge symmetry, are proportional to the $\alpha_i$'s, namely
\bq
&&\alpha_1 g_{\mu \nu} \equiv \int^{\Lambda}_k
\frac{g_{\mu\nu}}{k^2-m^2}-
2\int^{\Lambda}_k
\frac{k_{\mu}k_{\nu}}{(k^2-m^2)^2} \nonumber \\
&&=\int_k ^\Lambda \frac{\partial}{\partial k^\mu}
\left( \frac{k_ \nu}{(k^2-m^2)} \right),
\eq
\bq
&&\alpha_2 g_{\mu \nu} \equiv \int^{\Lambda}_k
\frac{g_{\mu\nu}}{(k^2-m^2)^2}-
4\int^{\Lambda}_k
\frac{k_{\mu}k_{\nu}}{(k^2-m^2)^3} \nonumber \\
&&=\int_k ^\Lambda \frac{\partial}{\partial k^\mu}
\left( \frac{k_ \nu}{(k^2-m^2)^2} \right)
\label{CR1}
\eq
and
\bq
&&\alpha_3 g_{\{\mu \nu \alpha \beta\}}   \equiv
g_{\{\mu \nu \alpha \beta \}}
\int^{\Lambda}_k
\frac{1}{(k^2-m^2)^2} \nonumber \\
&&-24\int^{\Lambda}_k
\frac{k_{\mu}k_{\nu}k_{\alpha}k_{\beta}}{(k^2-m^2)^4},
\label{CR2}
\eq
which means
\be
g_{\{\mu \nu \alpha \beta\}}(\alpha_3-\alpha_2)=\int_k^\Lambda \frac{\partial}{\partial k^\beta}
\left[ \frac{4k_\mu k_\nu k_\alpha}{(k^2-m^2)^3} \right].
\ee
In the equation above the limit $m^2 \to 0$ is supposed to be taken.
For the traditional spinorial QED, gauge invariance is obtained by setting these surface terms to zero.
Pauli-Villars Regularization, for example, is constructed to warrant gauge invariance,
and in this case the surface terms cancel out. The diagrammatic proof of
gauge invariance \cite{thooft} is based on the assumption that there exists a regularization which allows for shifts in the integration momenta.
Dimensional Regularization has been developed exactly with this characteristic. So, if
some technique works in the proper dimension of the theory, the preservation of the Ward identities depends on the
elimination of the surface terms by means of symmetry restoring counterterms. The exception takes place when
anomalies are involved. For situations like that, it is not possible to eliminate all the surface terms.

Now let us return to equation (\ref{QED}) and see how the result of \cite{altschul1} can be obtained. We turn our attention
to the second order contribution in the external vector $b_\mu$. We have
\bq
&&\frac 12\left(\Pi_{bb-}^{\mu \nu}+\Pi_{bb+}^{\mu \nu}\right) \nonumber \\
&& = -4\left\{ \left(b^2g^{\mu \nu} +2 b^\mu b^\nu \right)(\alpha_3-2\alpha_2) \right\}.
\eq
If one uses symmetric integration when calculating $\alpha_2$ and $\alpha_3$, such that $k^\mu k^\nu \to g^{\mu \nu}k^2/4$
and $k^\mu k^\nu k^\alpha k^\beta \to g^{\{\mu \nu \alpha \beta\}}k^4/24$, one obtains
\be
\alpha_2=\frac {i}{32\pi^2} \;\;\;\; \mbox{and} \;\;\;\; \alpha_3= \frac {5i}{96\pi^2},
\ee
so that
\be
\frac 12\left(\Pi_{bb-}^{\mu \nu}+\Pi_{bb+}^{\mu \nu}\right)=
\frac{i}{24 \pi^2}\left( b^2g^{\mu \nu} +2 b^\mu b^\nu \right).
\ee
It is the same result of \cite{altschul1} apart from some factors of $i$ and $e$ in the definition of the Feynman rules.
It is also interesting to see how the $b_\mu$ independent part of the photon self energy (the simple QED vacuum polarization
tensor) depend on the surface terms. From (\ref{QED}), we have
\bq
&&\Pi_{0}^{\mu \nu}=\Pi(p^2)(p^\mu p^\nu-p^2 g^{\mu \nu}) -4 \alpha_1 g^{\mu \nu}\nonumber \\
&&-\frac 43 \left\{ \alpha_2(p^\mu p^\nu-p^2 g^{\mu \nu}) \right. \nonumber \\
&& \left. + (2p^\mu p^\nu +p^2 g^{\mu \nu})(\alpha_3-2 \alpha_2)\right\}.
\eq
By examining the expression above, we conclude that if gauge invariance is broken by the
$\Pi_{bb\pm}^{\mu \nu}$ term, so it is also broken in the zeroth order, which is the
QED vacuum polarization tensor.
A simple calculation shows that any of the two possibilities for gauge invariance in
the zeroth and second order in $b_\mu$ photon self-energy,
the stronger one, $\alpha's=0$, or the weaker, $\alpha_1=0$ and $\alpha_3=2\alpha_2$, set to zero the first order dependence
of $\Pi_{\pm}^{\mu \nu}$ on $b_\mu$.

Hence, gauge invariance must be preserved in the full
amplitude. Note that this implies the complete disappearance of $b_\mu$ from $\Pi_{\pm}^{\mu \nu}$.
This means that the shifts $k \to k \pm b$ must be allowed independently in the two contributions, not necessarily
being a global shift.

However, even by enforcing gauge invariance, we still have the freedom of choice of the parameter $\alpha_2$. Now, we
have not considered yet the $\Pi^{\mu \nu}_{5\pm}$ parts. The two contributions are equal and we have
\be
\Pi^{\mu \nu}_{5}=\frac 12 \left( \Pi^{\mu \nu}_{5+}+\Pi^{\mu \nu}_{5-}\right)=\Pi^{\mu \nu}_{5+},
\ee
which, after Dirac algebra can be written as
\bq
&& \Pi^{\mu \nu}_{5}=-4ib_\beta \epsilon^{\mu \alpha \nu \beta}
\int_k^\Lambda \frac{(p+k)_\alpha}{(k+b)^2(k+p+b)^2} \nonumber \\
&&=-4ib_\beta \epsilon^{\mu \alpha \nu \beta}(p_\alpha I + I_\alpha),
\label{CFT}
\eq
with
\be
I,I_\alpha=\int_k^\Lambda \frac {1,k_\alpha}{(k+b)^2(k+p+b)^2}.
\ee
The results of these integrals by means of Implicit Regularization \cite{ir} are given by
\be
I=I_{log}(\lambda^2)-\frac{i}{16 \pi^2}\left[ \ln{ \left(-\frac{p^2}{\lambda^2}\right)}-2\right]
\ee
and
\bq
&&I_\alpha =-\frac{(p+2b)_\alpha}{2} \left\{ I_{log}(\lambda^2) \right. \nonumber \\
&& \left. -\frac{i}{16 \pi^2}\left[ \ln{ \left(-\frac{p^2}{\lambda^2}\right)}-2\right] -\alpha_2\right\},
\eq
where
\be
I_{log}(\lambda^2)=\int_k^\Lambda \frac{1}{(k^2-\lambda^2)^2}
\ee
is the remaining regularization dependent part and $\lambda^2$ is a mass parameter characteristic of
the procedure. Substituting these results in equation (\ref{CFT}), we get
\be
\Pi^{\mu \nu}_{5}=-4 i \alpha_2 p_\alpha b_\beta \epsilon^{\mu \nu \alpha \beta} \,\,\, \Rightarrow
\,\,\, \Delta c_\mu =2i \alpha_2 b_\mu.
\ee
We see that the coefficient of the CS-type generated term is proportional to the surface term $\alpha_2$.
It is the same parameter that could not be fixed on gauge invariance grounds. Of course, symmetric integration
will give us the traditional result
\be
\Delta c_\mu =-\frac {1}{16 \pi^2} b_\mu.
\ee
Nevertheless, we would like to argue that this result is ambiguous and regularization dependent. If we decide
to calculate all the surface terms by using symmetric integration, gauge invariance is violated even
in the zeroth order in $b$ (from eq. (\ref{QED}), transversality is violated in all orders in $b$ if symmetric
integration is applied to calculate the surface terms). This means violation of gauge symmetry in the simple QED. Moreover, the
non-perturbative functional calculus has shown us that the model must be gauge invariant.
Concerning the coefficient which was obtained in the functional calculation,  as discussed in
\cite{chung} and \cite{13}, there is an unavoidable ambiguity coming from the definition of the current
operator.

One comment on the meaning of Implicit Regularization is in order.
Although the regulator needs not to be specified, it
serves to obtain the following crucial features of IR: all infinities and the
differences of infinite integrals of the same degree of divergency (the surface terms)
occurring in a certain amplitude do not involve loop propagators that depend on
external momenta. Thus they decouple from the physical content and dynamics of the
amplitude. The latter is contained in strictly finite integrals, which are integrated without
restriction. As a consequence the regulator does not need to be a gauge invariant one
(an example is the simple cutoff), since the symmetry is restored by fixing the
surface terms (it corresponds to make use of symmetry restoring
counterterms). Nevertheless the regulator should not modify the dimension
of the integrals, or one would face the problems related to the dimension
specific theories. Keeping in mind that a regularization is implicit, all the integrals of
the same amplitude are treated on the same footing. In other words, the
same unregularized integrals will give us the same result when
regularized. The $\alpha_i$'s (see eq. (\ref{QED})) coming from them are the
same and are adjusted according with the symmetry.
On the other hand, since the adjustment of the $\alpha_i$'s corresponds
to using symmetry restoring counterterms, for different amplitudes they
can be fixed at different values. It is in this sense that we can say that
the same unregularized integrals can lead to different results, i.e. when they
stem from different amplitudes. This is not the case here. We are treating only one
amplitude, the vacuum polarization tensor of the model.

\section{The massive model}

In this section, we complement our analysis with the massive model. It is instructive to see that the
same conditions for preserving gauge symmetry are obtained.
First, we consider the complete photon self-energy, $\Pi_{\mu \nu}(p)$, which on gauge invariance grounds must be transverse:
\be
p^\mu \Pi_{\mu \nu}=0.
\ee
The modified fermionic propagator is given by
\be
S(k)=\frac{i}{\kbruto - m -\bbruto \gamma_5},
\ee
and so, we have
\bq
&&\Pi^{\mu \nu}= \nonumber \\
&&-\int_k^\Lambda \mbox {tr}\left\{ \gamma^\nu \frac{1}{\kbruto +\pbruto - m -\bbruto \gamma_5}
\gamma^\mu \frac{1}{\kbruto - m -\bbruto \gamma_5}\right\}. \nonumber \\
\eq
Now we perform the contraction with the external momentum $p$ and use the identity
$\pbruto= (\kbruto+\pbruto -m -\bbruto \gamma_5)- (\kbruto -m -\bbruto \gamma_5)$ and trace properties to obtain
\bq
&& p_\mu \Pi^{\mu \nu}= \int_k^\Lambda \mbox{tr} \left\{ \frac{1}{\kbruto - m -\bbruto \gamma_5}\gamma^\nu \right\} \nonumber \\
&& - \int_k^\Lambda \mbox{tr} \left\{ \frac{1}{\kbruto +\pbruto - m -\bbruto \gamma_5}\gamma^\nu \right\}.
\eq
And then it is clear from the equation above that gauge invariance is obeyed under the condition that
the shift $k \to k+p$ in the first integral is allowed (or $k \to k-p$ in the second one). Note that the
shift is not carried out in the amplitude as a whole, but only in one of the terms. In other words,
it is not a global shift. The same occurs if we are dealing with the traditional QED ($b=0$). Nevertheless,
the shift is carried out in the contracted amplitude $p_\mu \Pi^{\mu \nu}$ and this will be important
for the conclusions on the generation of the CS term.

Now, we consider the possibilities of violation of gauge symmetry, specifically the quadratic in $b$
contribution to the vacuum polarization tensor. For the modified fermion propagator,
it is possible to write an expansion, which reads
\bq
&&\frac{i}{\kbruto - m -\bbruto \gamma_5} = \sum_{n=0}^\infty \frac{i}{\kbruto - m}
\left\{ -i \bbruto \gamma_5 \frac{i}{\kbruto - m} \right\}^n \nonumber \\
&&= \sum_{n=0}^\infty S_n(k).
\eq
For the $b^2$ order, we have
\bq
&&\Pi^{\mu \nu}_{bb}=-\int_k^\Lambda \mbox{tr} \left\{ \gamma^\nu S_1(p+k)\gamma^\mu S_1(k) \right\} \nonumber \\
&& - \int_k^\Lambda \mbox{tr} \left\{ \gamma^\nu S_2(p+k)\gamma^\mu S_0(k) \right\} \nonumber \\
&&- \int_k^\Lambda \mbox{tr} \left\{ \gamma^\nu S_0(p+k)\gamma^\mu S_2(k) \right\}.
\eq
When the contraction with $p_\mu$ is performed and the identity $\pbruto= (\kbruto+\pbruto -m)- (\kbruto -m)$
is used, there remains only two terms:
\bq
&&p_\mu \Pi^{\mu \nu}_{bb}= \nonumber \\
&& -\int_k^\Lambda \mbox{tr} \left\{ \gamma^\nu \frac{1}{\kbruto +\pbruto -m} \bbruto \gamma_5
\frac{1}{\kbruto +\pbruto -m} \bbruto \gamma_5 \frac{1}{\kbruto +\pbruto -m} \right\} \nonumber \\
&& + \int_k^\Lambda \mbox{tr} \left\{ \gamma^\nu \frac{1}{\kbruto -m} \bbruto \gamma_5
\frac{1}{\kbruto -m} \bbruto \gamma_5 \frac{1}{\kbruto -m} \right\}.
\eq
The second term is null and the first differs from it by a surface term (they differ by a shift). So, all we have to do is to
identify the surface terms. They are easily identified by using, after the Dirac algebra has been performed,
the identity,
\bq
&& \frac {1}{(p+k)^2-m^2}=\frac{1}{(k^2-m^2)} \nonumber \\
&& -\frac{p^2+2p \cdot k}{(k^2-m^2) \left[(p+k)^2-m^2\right]},
\label{ident}
\eq
to separate the divergent (regularization dependent) terms. The surface terms come from differences between
integrals of the same degree of divergence. Using the definitions of equations (\ref{CR1}) and (\ref{CR2}), we get
\be
p_\mu \Pi^{\mu \nu}_{bb}= -4(\alpha_3-2\alpha_2)\left[b^2 p^\nu +2(b \cdot p)b^\nu \right],
\ee
and the same condition for transversality is achieved.

To complete our analysis, we look at the linear contribution in the external vector $b_\mu$. We have two contributions:
\bq
&& \Pi^{\mu \nu}_b= \int_k^\Lambda \mbox{tr} \left\{ \gamma^\nu S_0(p+k)\gamma^\mu S_1(k)\right\} \nonumber \\
&& +  \int_k^\Lambda \mbox{tr} \left\{ \gamma^\nu S_1(p+k)\gamma^\mu S_0(k)\right\}.
\eq
Using that
\be
S_1(k)=-iS_0(k)\bbruto \gamma_5 S_0(k),
\ee
we obtain
\bq
&& \Pi^{\mu \nu}_b= -i\int_k^\Lambda \mbox{tr} \left\{ \gamma^\nu S_0(p+k)\gamma^\mu S_0(k)\bbruto \gamma_5 S_0(k)\right\} \nonumber \\
&& -i  \int_k^\Lambda \mbox{tr} \left\{ \gamma^\nu S_0(k+p)\bbruto \gamma_5 S_0(k+p)\gamma^\mu S_0(k)\right\}.
\label{befshift}
\eq
We have already seen on gauge invariance grounds that a shift $k \to k \pm p$ must be allowed in $p_\mu \Pi^{\mu \nu}$,
not necessarily as a global shift. However, in this contracted amplitude the linear term in $b$ has already disappeared.
The reason is the presence of the antisymmetric Levi-C\`ivita tensor. So it does not say
anything about the surface term that emerges from a shift $k \to k-p$ in the first integral.
By performing this shift and using the fact that the integral is odd in $p$ and its antisymmetry under the
exchange $\mu \leftrightarrow \nu$, we see that we have two equal terms plus the surface term. This surface term is
easily identified. Then,
\bq
&&\Pi^{\mu \nu}_b= \nonumber \\
&& -2i b_\alpha \int_k^\Lambda \mbox{tr}
\left\{ \gamma^\nu S_0(k+p)\gamma^\alpha \gamma_5 S_0(k+p)\gamma^\mu S_0(k)\right\} \nonumber \\
&& +4i\alpha_2 b_\alpha p_\beta \epsilon^{\mu \nu \alpha \beta} \nonumber \\
&&\equiv -2i b_\alpha T^{\mu \nu \alpha} +4i\alpha_2 b_\alpha p_\beta \epsilon^{\mu \nu \alpha \beta}.
\label{aftshift}
\eq
We now follow the calculations of refs. \cite{scarpetal} and \cite{scarpetal2}. Carrying out the Dirac algebra, we get
\be
T^{\mu \nu \alpha}=\int_k^\Lambda \frac{{\cal N}^{\mu \nu \alpha}}{D},
\ee
with
\bq
&& {\cal N}^{\mu \nu \alpha}=-4\left\{ \left\{[(p+k)^2-m^2]
k_\beta-2m^2p_\beta \right\} \epsilon^{\mu \nu \alpha \beta} \right. \nonumber \\
&& \left. -2p_\sigma k_\beta k^\alpha \epsilon^{\mu \nu \sigma \beta} \right\}
\eq
and
\be
D=(k^2-m^2)[(p+k)^2-m^2]^2.
\ee
We can write
\bq
&& T^{\mu \nu \alpha}= -4 \left\{ (I_\beta -2m^2 p_\beta J)\epsilon^{\mu \nu \alpha \beta} \right. \nonumber \\
&& \left. -2 p_\sigma g^{\alpha \lambda} J_{\beta \lambda} \epsilon^{\mu \nu \sigma \beta}\right\},
\label{basic}
\eq
where
\be
I_\beta=\int_k^\Lambda \frac{k_\beta}{(k^2-m^2)[(p+k)^2-m^2]}
\ee
and
\be
J, J_{\beta \lambda}= \int_k^\Lambda \frac{1, k_\beta k_\lambda}{(k^2-m^2)[(p+k)^2-m^2]^2}.
\ee
The only finite integral is $J$. For the others a regularization method is needed.
Since we are going to analyze the regularization dependence of the generated term, we opt to
maintain the regularization implicit. So, we adopt here
the procedure of Implicit Regularization (IR) \cite{scarpetal},\cite{scarpetal2}, \cite{ir}, since it permits to let the evaluation
of divergent integrals to the end of the calculation.
The results of the divergent integrals are given by (see ref. \cite{scarpetal}):
\be
I_\beta=- \frac{p_\beta}{2} \left\{ I_{log}(m^2) -\frac{i}{16 \pi^2}Z_0(p^2,m^2)- \alpha_2 \right\}
\ee
and
\bq
&& J_{\beta \lambda}=\frac{g_{\beta \lambda}}{4} \left\{I_{log}(m^2) -\frac{i}{16 \pi^2}
Z_0(p^2,m^2)- \alpha_2 \right\}  \nonumber \\
&& + \mbox{terms in $p_\beta p_\lambda$},
\eq
where
\be
Z_0(p^2, m^2)= \int_0^1 dx \ln \left( \frac{p^2x(1-x)-m^ 2}{(-m^2)}\right).
\ee
For the finite one, we obtain
\be
J=\frac{i}{16 \pi^2} \int_0^1 dx \frac{(1-x)}{(p^2x(1-x)-m^ 2)}.
\ee
When these integrals are substituted into equation (\ref{basic}), only the term in $J$ survives:
\be
T^{\mu \nu \alpha}= 8m^2 J p_\beta \epsilon^{\mu \nu \alpha \beta}.
\ee
In the limit $p^2\to 0$, we have
\bq
&& T^{\mu \nu \alpha}\to \frac {i}{4 \pi^2}p_\beta \epsilon^{\mu \nu \alpha \beta} \nonumber \\
&& \Rightarrow   \;\;
\Pi_b^{\mu \nu} \to \frac {1}{2 \pi^2}b_\alpha p_\beta \epsilon^{\mu \nu \alpha \beta}\left(1+8i\pi^2\alpha_2\right).
\eq
This will give
\be
\Delta c_\mu=  \frac {1}{4 \pi^2}b_\mu\left(1+8i\pi^2\alpha_2\right).
\ee

Some comments are in order.
We can make explicit the regularization required to manipulate the divergent integrals.
Then, renormalization allows to remove the regulator in all the finite integrals and one
is left with the integrals which depend on the regularization, specifically
the basic divergence $I_{log}(m^2)$ and the surface term parameterized as $\alpha_2 g_{\mu \beta}$. We see that
$I_\beta$ and $J_{\mu \beta}$ do depend on the regularization to be used. Nevertheless, these ambiguous
terms cancel out exactly when the integrals are inserted into the amplitude.
Actually, the real ambiguity in all this calculation comes from the shift
carried out in one of the terms of eq. (\ref{befshift}) in order to obtain the expression (\ref{aftshift}).

The result of the massless model can be recovered if the limit $m^2 \to 0$ is taken before the
other one, $p^2 \to 0$. As before, gauge symmetry must be preserved, but it is not sufficient to
fix $\alpha_2$. The interesting result is that although the shift $k \to k\pm p$ must be allowed in
$p_\mu \Pi^{\mu \nu}$, this does not mean that all the surface terms must be set to zero. Actually,
they are constrained to respect a definite relation. Even in theories in which there are no parity violating
mathematical objects, like the $\gamma_5$ matrix, this freedom of fixing one of the surface terms occurs. However, in these cases
there is no loss of generality if they are put to zero.

\section{Summary and Discussion}

We have carried out an investigation of the modified Lorentz violating QED
with an axial term in the fermionic
sector. Particular attention was paid in the preservation or violation of
gauge symmetry of the model and its
relation with the radiative generation of a Chern-Simons-like term. If we
consider the action, we have shown
that the quantum corrections must preserve gauge symmetry. If it is
broken, then gauge invariance of the simple QED is also spoiled.
In order to come to this conclusion, the study of the massless model is very instructive, since in this
case the functional non-perturbative analysis (in $b$ and in the coupling constant), performed
in the beginning of section 2, exhibits clearly this feature.
When loop calculations are performed, the surface terms play an important role as parameters to be fixed.
Gauge invariance, in this case, requires the possibility of carrying out
some shifts. This fact does not restrict all the surface terms to be fixed null, although they are constrained
to obey a certain relation. As a conclusion, there is an unavoidable
ambiguity in the calculation of the coefficient of the radiatively generated Chern-Simons-like term.

However it has not been the main purpose of the present work to show that
gauge invariance does not fix the coefficient of the Chern-Simons
like term. This is only a residual conclusion.

It is obvious that the gauge invariance of
the action is not violated by the CS term, whatever its coefficient might
be. Our main focus is the quadratic term in b, the background that causes
Lorentz violation. Despite the existence of a vast bibliography concerning
the radiative generation of the Chern-Simons (CS) type term, only few
publications treat the quadratic piece.
Parity arguments have been considered by Altschul \cite{altschul1} to admit the possibility of violation of gauge invariance
by the quadratic term and the respective calculation in a specific scheme led
indeed to its violation. In further works \cite{altschul2}, \cite{altschul3}, this result is used to perform
other analysis. In ref. \cite{bonneau2}, the authors have performed the calculation of the quadratic term in $b$ using a
Pauli-Villars regularization and have obtained a gauge invariant result. However, they have used
a less general version of this regularization with the constraint that the gauge invariance
of the Lagrangian density is preserved. As a consequence, no Chern-Simons-like term is generated.
Therefore it was left no room for violation of gauge symmetry, and its preservation at $b^2$ order
brings no insight in what concerns the full amplitude. Moreover, the results of ref. \cite{altschul1}
has not been discussed.

In our present work we adopt the Implicit Regularization (IR) method to
perform a general analysis. For the massless case, for example, we analyze the complete one-loop
vacuum polarization tensor (all orders in b). We show that the conditions
to adjust the regularization dependent terms, in such a way that gauge
symmetry is preserved at $b^2$ order (recall that according to IR these
conditions are kept open until the end of the calculation), must be
exactly the same as for its preservation at zero-th order in b. In other
words, if gauge symmetry is spoiled at $b^2$ order, it is also spoiled in
QED. Furthermore, the functional non-perturbative analysis carried out in the first part of section 2,
based on ref. \cite{chung}, shows that the model is gauge invariant to all orders, as discussed in the paragraph
before equation (\ref{proj}).

Finally, we have also performed a critical discussion of some procedures
that are commonly adopted in the literature when shifts are done in
divergent integrals and the corresponding surface terms calculated in a "naive" way. We show
by comparing to the results obtained within IR, which is constructed to
yield symmetry preserving answers, that a naive implementation of surface terms may
lead to inconsistencies and violation of symmetries even in the simplest case (QED).

\vspace{0.5cm}

{\bf Acknowledgements}

Work supported in part by grants of Fundação para a Ci\^encia e Tecnologia, FEDER, POCI 2010, POCI/FP/63930/2005 and 81296/2007.

A. P. Ba\^eta Scarpelli and Marcos Sampaio are grateful to J. A. Helay\"el-Neto for very clarifying discussions.

\end{document}